\documentstyle[12pt]{report}
\def\lsim{\lower0.6ex\vbox{\hbox{$ \buildrel{\textstyle <}\over{\sim}\ $}}}
\def\rsim{\lower0.6ex\vbox{\hbox{$ \buildrel{\textstyle >}\over{\sim}\ $}}}
\begin{document}
\noindent
\phantom{DESY 90--???}           \hfill OSU-TA-22/94\\
 \vspace{ 2.5 cm}
\begin{center}
\begin{Large}
\begin{bf}
BIG BANG NUCLEOSYNTHESIS:\\
 CONSISTENCY OR CRISIS?\\
\end{bf} \end{Large}
\end{center}
 \vspace{.1cm}
\begin{center}
\begin{large}
GARY STEIGMAN\\
\end{large}
 \vspace{0.3cm}
Departments of Physics and Astronomy\\
The Ohio State University\\
174 West 18th Avenue\\
 Columbus, OH 43210, USA\\
\end{center}
 \vspace{1.8cm}
\centerline{\bf ABSTRACT}
\begin{quotation}
\noindent
The early hot, dense, expanding Universe was a primordial reactor in which
the light
nuclides D, $^3$He, $^4$He and $^7$Li were synthesized in astrophysically
interesting
abundances.  The challenge to the standard hot big bang model (Big Bang
Nucleosynthesis
$\equiv$ BBN) is the comparison between the observed and predicted
abundances, the
latter which depend only on the universal abundance of nucleons.  The
current status of
observations is reviewed and the inferred primordial abundances are used to
confront
BBN.  This comparison suggests consistency for BBN for a narrow range in
the nucleon
abundance but, looming on the horizon are some potential crises which will
be outlined.
\end{quotation}
\vfil
\newpage

\noindent
INTRODUCTION

Observations of an expanding Universe filled with black body radiation lead
naturally to
the inference that the early Universe was dense and hot and evolved through
an epoch in
which the entire Universe was a Primordial Nuclear Reactor.  During the
first $\sim$
thousand seconds ($\sim$ 20 minutes) the light elements D, $^3$He, $^4$He
 and
$^7$Li are synthesized in measurable abundances which range from $\sim 10
^{-10}$ (for
Li/H) to
$\sim 10^{-5}$ (for D/H and $^3$He/H) to $\sim 10^{-1}$ (for $^4$He/H) (for
a review and
references see Boesgaard \& Steigman 1985; for more recent results and
references see
Walker et al. 1991 (WSSOK)).  The predicted abundances depend on the
nucleon density,
conveniently measured by the ``nucleon abundance", the nucleon-to-photon
ratio $\eta
\equiv n_N/n_{\gamma} ~(\eta _{10} \equiv 10^{10} \eta$; for $T_{\gamma} $
= 2.726K,
$n_{\gamma} = 411 ~{\rm cm}^{-3}$).  Thus, BBN provides a test of the
consistency of the
hot big bang model and a probe of cosmology (e.g., of the universal density
of nucleons).
Specifically, is there a value (or a range of values) of $\eta$ such that
all the
predicted abundances are consistent with the inferred primordial abundances
derived from
the observational data?  Further, if there is consistency, is the inferred
nucleon
density (based on processes which occurred during the first $\sim 10^3$
sec. of the
Universe) consistent with that observed at present (when the Universe is
$\sim $10 Gyr
old)?

According to WSSOK, both questions are answered in the affirmative with
$2.8~ \lsim~ \eta
_{10}~ \lsim 4.0$.  The nucleon density parameter ($\Omega _N \equiv \rho
_N/\rho
_{CRIT}$) is related to the nucleon abundance and the Hubble parameter
($h_{50} \equiv
H_0/50 kms ^{-1} Mpc ^{-1}$) by,

$$\Omega _N h{_{50} ^2} = 0.015 ~\eta _{10}, \eqno (1)$$
(for $T_{\gamma} = 2.726 \pm 0.010 (2 \sigma)$, the coefficient in (1)
varies from
0.0145 to 0.0148).  Thus, for $2.8~ \lsim ~\eta _{10} ~\lsim ~4.0,~ 0.04
{}~\lsim~ \Omega _N
h{^2 _{50}} ~\lsim ~0.06$, which leads to the conclusion that there are
dark baryons
$(\Omega _N > \Omega _{LUM}$) but, not all dark matter is baryonic ($\Omega
_N < \Omega
_{DYN}$).

The physics of BBN is, by now, well understood; for overviews see Boesgaard
\& Steigman
(1985) and Smith, Kawano \& Malaney (1993).  It is, however, worth
emphasizing that
Primordial Alchemy is conventional physics.  For example, the timescales
are long $(
\sim 10^{-1} - 10^3$ sec.) and the temperatures (thermal energies) are low
(kT$ \sim
10$ keV - 1 MeV) .  Although the early Universe is dense, it is dilute on
the scale of
nuclear physics during the epoch of BBN.  For example, for T $\lsim $1/2
MeV, the
internucleon separation is $\rsim 10^6$ fermis.  Thus, collective and/or
many body
effects are entirely negligible.  The nucleon reaction network is very
limited
(effectively, $A \le 7$) and simple.  More importantly, the cross sections
are measured
at lab energies comparable to the thermal energies during BBN.  Thus, in
stark contrast
to stellar nucleosynthesis (where $kT_{\star} \ll E_{1ab}$), large and
uncertain
extrapolations are not required.  Thus, for fixed $\eta$, the BBN predicted
abundances
of D, $^3$He and $^7$Li are known to better than $\sim$ 20 \% and the
$^4$He mass
fraction ($Y_{BBN}$) is known to $\vert \delta Y_{BBN} \vert ~\lsim~6
\times 10 ^{_4}$
(Thomas et al. 1994).

Since the BBN abundances of D, $^3$He and $^7$Li vary noticeably with
$\eta$, those
nuclides serve as ``baryometers", leading to constraining lower and upper
bounds to
$\eta$ (e.g., WSSOK).  $Y_{BBN}$ varies little $(\sim$ logarithmically)
with $\eta$ and,
thus, serves as the key to testing the consistency of BBN.  In the next
sections we
first survey the observational data on D, $^3$He and $^7$Li and derive
bounds to their
primordial abundances.  Next, the predicted and inferred primordial
abundances are
compared to test for consistency and to bound $\eta$.  Then, the $^4$He
abundance is
studied for consistency -- or crisis.  Finally, the health of BBN is
assessed and possible
crises are outlined.

\noindent DEUTERIUM

BBN is the only source of astrophysical deuterium.  Whenever cycled through
stars, D is
destroyed (burned to $^3$He, even during pre-main sequence evolution).
Thus, the mass
fraction ($X_2$) of primordial deuterium is no smaller than that observed
anywhere in
the Universe:  $X_{2P} \ge X_{2OBS}$.

As with all of the light elements, there is both bad news and good news.
The bad news is
that, at least until recently (possibly!), deuterium has been observed only
locally (in
the interstellar medium (ISM) and the solar system).  The good news is that
the data is
accurate.

Geiss (1993) has reanalyzed the solar system D and $^3$He data.  Using
Geiss' results,
Steigman \& Tosi (1994) find

$$X_{2\odot} = 3.6 \pm 1.3 \times 10^{-5} . \eqno (2)$$
Using older Copernicus and IUE data, along with newer HST data (Linsky et
al. 1993),
Steigman \& Tosi (1994) have noted that over a range of two orders of
magnitude in HI column
density, (D/H)$_{ISM}$ is constant at a value of $1.6 \pm 0.2 \times 10^{-
5}$.  To
determine $X_{2 ISM}$ requires knowledge of the H mass fraction in the ISM,
for $Y_{ISM}
\approx 0.28 \pm 0.02 ~{\rm and}~ Z_{ISM} \approx 0.02, ~ X_{ISM} \approx
0.70 \pm 0.02$
and,

$$X_{2ISM} = 2.2 \pm 0.3 \times 10^{-5}. \eqno (3)$$
It is expected that $X_2$ should have decreased in the 4.6 Gyr between the
formation of
the solar system and the present (although as Steigman and Tosi (1992)
show, the
decrease may be small).  The data are marginally consistent with this
expectation:
$X_{2\odot}/X_{2ISM} = 1.6 \pm 0.6$.

A lower bound to $X_{2OBS}$ leads to a lower bound to $X_{2P}$ which, in
turn, leads to
an upper bound to $\eta$.  For $X_{2P} \ge X_{2ISM} \ge 1.7 \times 10^{-5}
(2 \sigma)$,

$$D: \hspace{.1in} \eta _{10} \le 9.0 \eqno (4)$$

\eject
\noindent HELIUM-3

When deuterium is cycled through stars it is burned to $^3$He.  $^3$He
burns at a higher
temperature than D so that $^3$He survives in the cooler, outer layers of
stars.
Furthermore, since hydrogen burning is incomplete in low mass stars, such
stars are net
sources of $^3$He.  Thus, any primordial $^3$He is modified by the
competition between
stellar production and destruction and, therefore, a detailed evolution
model -- with
its attendant uncertainties -- is needed to relate the observed and BBN
abundances
(Steigman \& Tosi 1992).  However, since {\it all} stars do burn D to
$^3$He and, {\it
some} $^3$He does survive stellar processing, the primordial D + $^3$He may
be bounded
by the observed D and $^3$He (Yang et al. 1984 (YTSSO); Dearborn, Schramm
\& Steigman
1986).  The YTSSO analysis, which has recently been updated (Steigman \&
Tosi 1994), is
``generic" in the sense that it should be consistent with any specific
model for
Galactic chemical evolution.  Its predictions do, however, depend on one
model
specific parameter $g_3$, the ``effective" survival fraction of $^3$He.

Since the deuterium observations have already been used to bound the
primordial D mass
fraction from below, here we are interested in using the solar system
observations of D
and $^3$He to bound $X_{2P}$ from above.  If any net stellar production of
$^3$He is
neglected (so that $^3$He only increases by burning D and decreases by
stellar
destruction), it can be shown that (YTSSO; Steigman \& Tosi 1994)

$$ X_{2P} < X{^{MAX} _{2P}} = \bigg [ 1 - {1 \over g_3} \bigg ({y_3 \over
y_{23}} \bigg
)_P \bigg ] X_{2 \odot} + {2/3 \over g_3} \bigg ( {y_2 \over y_{23} } \bigg
) _P~ X_{3
\odot}. \eqno (5)$$
In (5), the primordial D and $^3$He abundances (by number) are $y_{2P} =
(D/H)_P$ and
$y_{3P}$ = ($^3$He/H)$_P$; $y_{23P} = y _{2P} + y _{3P}$; $g_3$ is the
``effective"
survival fraction of $^3$He (which is model dependent).  It can be seen
from (5) that the
higher/lower the primordial/solar system $^3$He abundances, the more
restrictive the
upper bound on primordial deuterium.

Of course, since primordial abundances appear on both sides of eq. 5, care
must be
excersized in finding the bound.  One approach is to evaluate both sides of
(5) using
the {\it predicted} abundances as a function of $\eta$, identifying those
values of
$\eta$ for which the inequality is satisfied (Steigman \& Tosi 1994).
Alternatively, the
inequality can be further relaxed by entirely neglecting {\it any}
primordial $^3$He.
Since $y_{3P} > 0$, we may write,

$$X_{2P} < X{^{MAX} _{2P}} < \big ( X{^{MAX} _{2P}} \big)^0 = X_{2 \odot} +
\Bigg ( {2/3
\over g_3 }\Bigg ) X_{3 \odot}. \eqno (6)$$
The inequality in (6) may be further reinforced to relate $y_{2P}$ to $y_{2
\odot}$ and
$ y_{3 \odot}$ since the hydrogen mass fraction always decreases from its
primordial
value $(X _{H \odot} < X_{HP}$),

$$ y_{2P} < \big ( y{ ^{MAX} _{2P}} \big)^0 < y_{2 \odot} + g{^{-1} _3}
y_{3 \odot}.
\eqno (7)$$

Using the Geiss (1993) solar system abundances and $g_3 > 1/4$ (Dearborn,
Schramm \&
Steigman 1986), Steigman \& Tosi (1994) find $X_{2P} < 11 \times 10 ^{-5}
{}~(y_{2P} < 7.4
\times 10^{-5})$ which leads to a {\it lower} bound to $\eta$,

$$ D + ~^3{\rm He}: \hspace {0.1in} \eta _{10} ~\rsim ~3.1 \eqno (8)$$
Note that if the more restrictive survival fraction $g_3 > 1/2$ (Steigman
\& Tosi 1992)
is used, we would infer $X_{2P} < 7 \times 10^{-5}$ and $\eta _{10}~ \rsim
4$.  It
should also be noted that $2 \sigma$ upper bounds to $X_{2 \odot}$ and
$X_{3 \odot}$ are
used in reaching these conclusions.

To summarize the progress so far,  solar system and interstellar
observations of D
and $^3$He have permitted us to bound primordial deuterium from above and
below $(1.6~
\lsim ~10^5 X_{2P}~ \lsim 11)$ which leads to consistent upper and lower
bounds on $\eta
{}~(3.1~\lsim ~\eta _{10} ~\lsim 9.0)$.  Next, we turn to the first
consistency test of BBN
by considering lithium-7.

\noindent LITHIUM-7

As with the other light nuclides, the status of lithium observations has
good news and
bad news.  The good news is that lithium is observed, with relatively good
statistical
accuracy, in dozens and dozens of stars of varying metallicity, mass (or
temperature),
evolutionary stage, population, etc.  Among the bad news, these stars are
all in the
Galaxy and, therefore, provide a sample which is not necessarily universal.
More
serious, however, are the essential corrections which are required to go
from the
observed surface abundances to their unmodified (by stellar evolution)
prestellar values
and, to account for the production/destruction of lithium in the course of
Galactic
chemical evolution.

The overwhelming influence of stellar evolution on the stellar surface
lithium
abundance is reflected in the enormous range of observed values in
Population I stars.
The Sun is a case in point.  Whereas the meteoritic abundance of lithium is
$\sim 2
\times 10^{-9} ([{\rm Li}] \equiv 12 + \log ({\rm Li/H}) = 3.31)$, the
solar photosphere
abundance is smaller by some two orders of magnitude (Grevesse \& Anders
1989).  There
is, however, evidence for a maximum PopI lithium abundance as inferred from
observations
of the warmest stars in young open clusters (Balachandran 1994),
[Li]$_{PopI} = 3.2 \pm
0.2 (2
\sigma)$.  And, further, there is evidence (e.g., Beckman, Robolo \& Molaro
1986) that
this maximum decreases with decreasing metalicity until, for  [Fe/H] $\lsim
-1.3$, the
``Spite Plateau" is reached.

The Spites' discovery (Spite \& Spite 1982a,b), subsequently confirmed by
many
observations (e.g., see WSSOK for an overview and references and, see
Thorburn 1994 for
the latest observations), is that the warmest $(T ~\rsim ~5700$K), most
metal-poor stars
 ([Fe/H] $\lsim -1.3)$ have, with remarkably few exceptions, the {\it same}
lithium
abundance:   [Li]$_{PopII} \approx 2.1$ (WSSOK; the values from Thorburn
(1994) are
systematically higher by $\sim$ 0.2 dex).  The value of the Spites'
discovery cannot be
overestimated but, too, caution is advised.  On the one hand, the
``plateau" in Fe/H
(or, where available, in oxygen abundance) suggests that  [Li]$_{PopII}$
may provide an
estimate of the primordial abundance free from a (significant) correction
for Galactic
chemical evolution.  On the other hand, the temperature plateau suggests
that, ``what
you see is what you get".  That is, the surface abundances of lithium in
the warmest
PopII stars provide a fair sample of the lithium abundances in the gas out
of which
those stars formed.  If, indeed,  [Li]$_P \approx [{\rm Li}]_{PopII}
\approx 2.1 \pm 0.2$
(the uncertainty is mainly systematic, the statistical uncertainties are
much smaller
(WSSOK)), then BBN is constrained significantly; for  (Li/H)$_{BBN}~ \lsim
2 \times
10^{-10},~ 1.6 ~\lsim ~\eta _{10}~ \lsim~ 4.0$.  However, analysis of
Thorburn's (1994)
extensive data set raises questions about the {\it flatness} of the lithium
temperature/metallicity plateaus.

Furthermore, it is not clear that corrections for chemical evolution are
entirely
negligible, even for the very old, very metal-poor PopII stars.  Lithium-7
(as well as
$^6$Li) may be produced by $\alpha - \alpha$ fusion reactions in Cosmic Ray
Nucleosynthesis (CRN; Steigman \& Walker 1992) as well as by the more
familiar
spallation reactions of p and $\alpha$ on CNO nuclei.  Since the spallation
reactions
require CNO targets (and/or projectiles) whereas the fusion reactions can
utilize
primordial $^4$He, CRN lithium production has a component which is
shallower in its
metallicity dependence than that of Be and/or B which are only synthesized
in
spallation reactions.  Thus, if  (Be/H)$_{PopII} \sim({\rm
Fe/H})^{\alpha},~ \Delta
(^7{\rm Li/H})_{\alpha \alpha} \sim ({\rm Fe/H})^{\alpha -1}$ and, since
current data
(Gilmore et al. 1992; Boesgaard \& King 1993) suggests $\alpha \approx 1, ~
(\Delta
y_7)_{\alpha
\alpha}$ should be nearly independent of metallicity and, so, will mimic a
primordial
component $(y_7 \equiv  ^7$Li/H).  Thus, even neglecting any early (PopII)
stellar
production/destruction of $^7$Li, the BBN and observed PopII lithium
abundances are, in
general related by,

$$ y_{7OBS} = f_7 (y_{7BBN} + (\Delta y_7)_{CRN}), \eqno (9)$$
where $f_7 (\le 1)$ is the stellar surface destruction/dilution factor for
$^7$Li.
Although ``standard" (i.e., nonrotating) models for the warmest PopII stars
suggest
$f_7 \approx 1$ (Chaboyer et al. 1992), models with rotation may permit a
significant
correction $(f_7 ~\rsim 0.1 - 0.2$; Pinsonneault, Deliyannis \& Demarque
1992; Charbonel
 \& Vauclair 1992).  The
observations of the much more fragile $^6$Li in two PopII stars (Smith,
Lambert \& Nissen
1992; Hobbs \& Thorburn 1994) suggests that $f_7
\approx 1$ but this important issue remains unresolved at present.  Thus,
although the
PopII stellar data appears consistent with  [Li]$_{BBN}~ \lsim 2.3$, it is
unclear that
the much higher bound  [Li]$_{BBN} ~\lsim 3.0$ (Pinsonneault, Deliyannis \&
Demarque 1992)
can be entirely excluded.

Fortunately, another -- independent -- path to primordial lithium exists.
Lithium has
been observed in the ISM of the Galaxy (Hobbs 1984; White 1986) and,
searched for in
the ISM of the LMC (in front of SN87A; Baade et al. 1991).  The
interstellar data has
assets and liabilities of its own which, however, are {\it different} from
those of the
stellar data.  Among the liabilities is a large and uncertain ionization
correction
since LiI is observed but most ISM $^7$Li is LiII.  Another problem is the
correction
for lithium removed from the gas phase of the ISM (where it is observed) by
grains
and/or molecules (where it is unobserved).  Steigman (1994a) has proposed
avoiding
these obstacles by comparing lithium to potasium (which shares the
ionization/depletion
problems with lithium) and evaluating the {\it relative} abundances (Li/K
rather than
Li/H).  Comparing Galactic ( [Fe/H] $\approx 0)$ Li/K with the absence of
Li and the
presence of K in the LMC  ([Fe/H]$_{LMC} \approx -0.3)$, Steigman (1994a)
has concluded
that  (Li/K)$_{LMC} ~\lsim 1/2 ({\rm Li/K})_{GAL}$. Since potassium has no
primordial
component, this bound can be used to derive an upper bound to primordial
lithium
(Steigman 1994a):  [Li]$_P ~\lsim 2.3 - 2.8$.  Thus, although it appears
that the Spite
Plateau bound   [Li]$_{BBN} ~\lsim 2.3$ is supported, a higher value cannot
be entirely
excluded.  Here, in the absence of evidence to the contrary, I will use the
above bound
((Li/H)$_{BBN} ~\lsim 2 \times 10 ^{-10})$ to constrain $\eta$,

$$^7{\rm Li}:  1.6~ \lsim \eta _{10}~ \lsim 4.0. \eqno (10)$$

\noindent CONSISTENCY AMONG D, $^3$He \& $^7$Li?

Before moving on to the keystone of BBN, helium-4, it is useful to pause at
this point
to consolidate the progress thus far.  Solar system and interstellar
observations of D
and $^3$He have been employed to set lower and upper bounds to primordial
deuterium
$(1.6 ~\lsim 10^5 X_{2P}~ \lsim 11)$ which result in bounds on the nucleon
abundance
$(3.1 ~\lsim \eta _{10} ~\lsim 9.0)$. PopII and ISM observations of lithium
are consistent
with an upper bound on primordial lithium which may be as small as
(Li/H)$_P~ \lsim 2
\times 10 ^{-10}$ but, which could also be consistent with a larger value
(Li/H)$_P~
\lsim 6 - 8 \times 10 ^{-10}$.  Utilizing the more restrictive lithium
bound,
consistency among the BBN predicted abundances is achieved provided that
$\eta$ is
restricted to a relatively narrow range,

$$D, ~^3{\rm He}, ~^7{\rm Li}: \hspace {0.1in} 3.1~ \lsim ~\eta _{10}~
\lsim ~ 4.0. \eqno
(11)$$ From (1) it follows that the present density in nucleons is
similarly restricted,

$$ 0.045 ~\lsim ~\Omega _N h{^2 _{50}} ~\lsim ~ 0.059 \eqno (12)$$
which, for $40 \le H_0 \le 100 kms^{-1} Mpc ^{-1}$, corresponds to, $0.011
{}~\lsim \Omega
_N ~\lsim 0.093$. The lower bound $\Omega _N ~\rsim 0.01$ exceeds the
estimate of the
mass associated with ``luminous" matter, suggesting the presence of
Baryonic Dark
Matter, while the restrictive upper bound $\Omega _N ~\lsim 0.09$ is strong
evidence for
the existence of Non-Baryonic Dark Matter.

\noindent HELIUM-4

The good news about $^4$He is that it is ubiquitous and can be seen
everywhere in the
Universe.  And, since its abundance is large, its value can be determined
with high
statistical accuracy.  The bad news is that the path from observations to
abundances to
primordial helium is strewn with corrections which are accompanied by
potentially
large systematic uncertainties.

As stars burn, hydrogen is consumed producing $^4$He which is returned to
the galactic
pool out of which subsequent generations of stars form.  Thus, any observed
abundances
must be corrected for the $^4$He enhancement from the debris of earlier
generations of
stars.  To minimize this correction and its attendant uncertainties, the
most valuable
observational data is that from the low metallicity, extragalactic HII
regions (e.g.,
Pagel et al. 1992).  It is the emission lines from the recombination of
$^4$He$^+$ and
$^4$He$^{++}$ (as well as H$^+$) which are observed from these regions.
Since neutral
helium (in the zone of ionized hydrogen) is unobserved, its correction --
which carries
with it systematic uncertainties -- is minimized by restricting attention
to the
hottest, highest excitation regions (Pagel et al. 1992) where the
correction may be
negligible (or, even, negative in the sense that HII regions ionized by
very hot --
metal-poor -- stars may have HeII zones larger in extent than the HII
zones).  Finally,
to benefit from the high statistical accuracy of the observational data,
corrections
for collisional excitation, radiation trapping and destruction by dust,
etc. must be
considered.

The best (i.e., most coherent) data set of Pagel et al. (1992) has recently
been
supplemented (Skillman et al. 1993) by the addition of $\sim$ a dozen
 very low
metallicity HII regions.  Olive and Steigman (1994) have analyzed this
data; there are
some four dozen HII regions whose oxygen abundances extend down to $\sim$
1/50 solar
and whose nitrogen abundances go down to $\sim 1\%$ of solar.  For this
data Olive and
Steigman (1994) find that an extrapolation to zero metallicity yields,

$$ Y_P = 0.232 \pm 0.003, \eqno (13)$$
where the uncertainty is a $1 \sigma$ {\it statistical} uncertainty.  Thus,
at $2
\sigma$, $Y_{BBN}~ \lsim 0.238$.  It is difficult to estimate the possible
{\it
systematic} uncertainty; Pagel (1993), WSSOK, and Olive \& Steigman (1994)
suggest $
\pm 0.005 ~({\rm i.e.}, \sim 2 \%)$.  If so, the upper bound may be relaxed
to $Y_{BBN}~
\lsim 0.243$ which, as will be seen shortly, may be crucial.

The BBN predicted $^4$He mass fraction is known to high accuracy (as a
function of
$\eta$). For the standard case of three light neutrinos $(N_{\nu} = 3)$ and
a neutron
lifetime in the range $\tau _n = 889 \pm 4 (2 \sigma)$ sec, the bounds from
observation
$Y_{BBN} \le 0.238 (0.243)$ require $\eta _{10} \le 2.5 (3.9)$.  Here, we
have the first
serious crisis confronting BBN! Unless systematic corrections increase the
primordial
abundance of helium inferred from the observational data, the upper bound
on $\eta$
from $^4$He is exceeded by the lower bound on $\eta$ from D (and $^3$He).
With,
however, allowance for a possible $\sim 2\%$ uncertainty, consistency is
maintained.
Thus, for D, $^3$He, $^7$Li and $Y_{BBN} \le 0.243$,

$$ 3.1 \le \eta _{10} \le 3.9. \eqno (14)$$
Of course, the upper bound to $\eta$ from $^4$He will reflect the
uncertainty in the
obserational bound to $Y_P$. For $\eta _{10} \sim 4,~ \Delta Y_{BBN}
\approx 0.012
(\Delta \eta / \eta$) so that an uncertainty of 0.003 in Y corresponds to a
25\%
uncertainty in $\eta (\Delta \eta _{10} \approx \pm 1$).

The importance of $^4$He is that the predicted primordial abundance is
robust --
relatively insensitive to $\eta$ and, as a function of $\eta$, accurately
calculated (to
better than $\pm 0.001$). And, being abundant, $^4$He is observable
throughout the
Universe and, systematic uncertainties aside, the derived abundance is
known to high
statistical accuracy $(\lsim \pm 0.003)$. Thus, $^4$He is the keystone to
testing the
consistency of BBN.

\eject
\noindent A HELIUM-4 CRISIS?

Solar system data on D and $^3$He, along with a ``generic" model for
galactic
evolution (Steigman \& Tosi 1994) leads to a {\it lower} bound to $\eta ~ (
\eta _{10}
{}~\rsim 3.1)$ and, therefore, to a {\it lower} bound to the predicted BBN
abundance of
$^4$He; for $N_{\nu} = 3,~ \tau _N \ge 885$~sec and $\eta _{10} \ge 3.1,~
Y_{BBN} \ge
0.241$. In contrast, accounting only for statistical uncertainties, $Y_P
\le 0.238$ (at
$ 2 \sigma$; Olive \& Steigman 1994). Thus, the issue of whether or not
this is a
crisis for BBN hinges on whether or not $Y_P$ is known to three significant
figures.
Allowance for a possible, modest $(\sim 2\%$), systematic uncertainty of
order 0.005 would
transform this potential crisis to consistency.

\noindent A DEUTERIUM CRISIS?

Recently, two groups have independently reported the possible detection of
extragalactic
deuterium in the spectrum of a high z (redshift), low Z (metallicity) QSO
absorption
system (Songaila et al. 1994; Carswell et al. 1994). If, indeed, the
absorption is due
to deuterium, the inferred abundance is surprisingly high: $D/H \approx 19 -
25 \times
10^{-5}$. This high abundance -- an order of magnitude larger than the pre-
solar or ISM
values -- poses no problem for cosmology in the sense that for $(D/
H)_{BBN} \sim
2 \times 10^{-4},~ \eta _{10} \sim 1.5 ~{\rm and}~ Y_{BBN} \sim 0.23 ~{\rm
and}~
(^7{\rm Li/H})_{BBN} \sim 2 \times 10 ^{-10}$, which are in excellent
agreement with the
observational data.  If, indeed, $\eta _{10} \sim 1.5$, then $\Omega _N
h{^2 _{50}}
\sim 0.022$,  reinforcing the argument for non-baryonic dark matter (for
$H_0 ~\rsim
40 kms ^{-1} Mpc ^{-1},~ \Omega _N ~\lsim 0.034$)

But, such a high primordial abundance does pose a serious challenge to our
understanding of the stellar and galactic evolution of helium-3. The issue
is that if
$\sim 90 \%$ of primordial deuterium has been destroyed prior to the
formation of the
solar system, then the solar nebula abundance of $^3$He should be much
larger than
observed (Steigman 1994b) since D burns to $^3$He and some $^3$He survives.
Earlier, we
have used the solar system data to infer a primordial bound $y_{2P} < 7.4
\times 10
^{-5}~({\rm for} ~ g_3 \ge 1/4$). A primordial abundance as large as $\sim
2 \times
10^{-4}$ would require much more efficient stellar destruction of $^3$He
($g_3 ~\lsim
0.09$).

It is, however, possible that the observed absorption feature is not due to
high z, low Z
deuterium at all but, rather, to a hydrogen interloper (Steigman 1994b).
That is, the
absorption may be from a very small cloud of neutral hydrogen whose
velocity is shifted
from that of the main absorber by just the ``right" amount so that it
mimics deuterium
absorption.  As Carswell et al. (1994) note, the probability for such an
accidental
coincidence is not negligible ($\sim 15 \%$).  This possibility can only be
resolved
statistically when there are other candidate D-absorbers. Data from Keck
and the HST is
eagerly anticipated.

\noindent THE X-RAY CLUSTER CRISIS?

This overview concludes with a glimpse of yet another potential crisis for
BBN.  Large
clusters of galaxies are expected to provide a ``fair sample" of the
Universe in the
sense that, up to factors not much different from unity, the baryon
fraction in clusters
should be the same as the universal baryon fraction.

$$ f_B = \Omega _B/\Omega \approx (M_B/M_{TOT})_{Clusters} \eqno (15)$$
In (15) the baryon density parameter $\Omega _B$ is another name for what
we have been
calling the nucleon density parameter $\Omega _N$ and $\Omega$ is the ratio
of the total
density to the critical density.  For x-ray clusters $M_B$ is dominated by
the mass in
hot, x-ray emitting, intercluster gas $(M_B \approx M_{HG} + M_{GAL} ~\rsim
M_{HG}$) so
that,

$$ \Omega ~\lsim \Omega _{BBN} / f_{HG}, \eqno (16)$$
where $\Omega _{BBN} = 0.015 \eta _{10} h{_{50} ^{-2}}$ and $f_{HG}$ is the
hot gas
fraction in x-ray clusters. Since $M_{HG}$ and $M_{TOT}$ scale differently
with the
distance to the cluster, $f_{HG}$ depends on the choice of Hubble parameter
(e.g.,
Steigman 1985): $f_{HG} = A_{50} h{_{50} ^{-3/2}}$. Thus, (16) may be
written as,

$$\Omega h{^{1/2} _{50}} ~\lsim 0.6 \Bigg ( {0.10 \over A_{50}} \Bigg )
\Bigg( {\eta _{10}
\over 4.0} \Bigg), \eqno (17)$$
where x-ray observations yield $A_{50}$.

The x-ray cluster crisis was perhaps first noted by White et al. (1993) for
Coma where:
$A_{50} (Coma) = 0.14 \pm 0.04$. For $A_{50} ~\rsim 0.10 ~{\rm and} ~\eta
_{10} ~\lsim
4.0,~ \Omega =1$ requires $H_0 ~\lsim 18 kms^{-1} Mpc^{-1}$! Thus, either
$\Omega < 1$ or
the BBN upper bound to $\eta$ is wrong.  Further x-ray data, however, makes
this latter
choice less likely. White et al. (1994) find for Abell 478, $A_{50}(A478) =
0.28 \pm
0.01$, a result supported by White \& Fabian's (1994) survey of 19 x-ray
clusters which
finds, at an Abell radius of $ \sim 3 h{^{-1} _{50}} Mpc,
A_{50} \approx 0.24.  ~{\rm For}\ A_{50} \approx 0.24,\ \Omega h {^{1/2}
_{50}} ~\lsim ~
\eta _{10} /16$, strongly hinting at $\Omega <1$. For $\Omega = 1$ and any
sensible choice
of
$H_0,~
\eta _{10}$ would have to be so large as to violate -- separately -- the
observational
bounds on D,
$^4$He, and $^7$Li. The x-ray cluster crisis -- if real --  is a crisis for
$\Omega =1$
but, not for BBN.

BBN is alive and well and the healthy confrontation of theory with
observation continues.

 \noindent REFERENCES

\begin{itemize}
\item Baade, D., Cristiani, S., Lanz, T., Malaney, R.A., Sahu, K. C. \&
Vladilo, G. 1991,
{\it A \& A}, {\bf 251}, 253
\item Balachandran, S. 1994, Preprint (Submitted to {\it ApJ\/})
\item Beckman, J. E., Rebolo, R. \& Molaro, P. 1986, ``Advances in Nuclear
Astrophysics"
(eds. E. Vangioni-Flam, J. Audouze, M. Cass\'e, J. P. Chi\`eze \& J.
TranThanh Van; Editions
Fronti\'eres) p. 29
\item Boesgaard, A. M. \& Steigman, G. 1985, {\it Ann. Rev. Astron.
Astrophys.} {\bf 23},
319
\item Boesgaard, A. M. \& King, J. R. 1993, {\it AJ\/} {\bf 106}, 2309
\item Carswell, R. F., Rauch, M., Weymann, R. J., Cooke, A. J. \& Webb, J.
K. 1994, {\it
MNRAS\/} {\bf 268}, L1
\item Chaboyer, B., Deliyannis, C. P., Demarque, P., Pinsonneault, M. H. \&
Sarajedini, A. 1992, {\it ApJ\/}, {\bf 388}, 372
\item Charbonnel, C. \& Vauclair, S. 1992, {\it A \& A}, {\bf 265}, 55
\item Geiss, J. 1993 ``Origin and Evolution of the Elements" (eds. N.
Prantzos, E.
Vangioni-Flam
\& M. Cass\'e; Cambridge Univ. Press) p. 89
\item Gilmore, G., Gustafsson, B., Edvardsson, B. \& Nissen, P. E. 1992,
{\it Nature}, {\bf
357}, 379
\item Grevesse, N. \& Anders, E. 1989, in {\it ATP Conf. Proc. 183},
``Cosmic Abundances of
Matter", (ed. C. J. Waddington; AIP), p.1
\item Hobbs, L. M. 1984, {\it ApJ\/} {\bf 286}, 252
\item Hobbs, L. M. \& Thorburn, J. A. 1994, {\it ApJ\/} {\bf 428}, L25
\item Linsky, J. L.,
Brown, A., Gayley, K., Diplas, A., Savage, B. D., Ayres, T. R., Landsman,
W., Shore, S. W.
\& Heap, S. 1993, {\it ApJ\/} {\bf 402}, 694
\item Olive, K. A. \& Steigman, G. 1994, {\it ApJS\/} In Press
\item Pagel, B. E. J., Simonson, E. A., Terlevich, R. J. \& Edmunds, M. G.
1992, {\it
MNRAS\/} {\bf 255}, 325
\item Pagel, B. E. J. 1993, {\it Proc. Nat. Acad. Sci.} {\bf 90}, 4789
\item  Pinsonneault, M. H., Deliyannis, C. P. \& Demarque, P. 1992, {\it
ApJS\/} {\bf 78},
179
\item Skillman, E. D., Terlevich, R. J., Terlevich, E., Kennicutt, R. C. \&
Garnett, D. R.
1993, {\it Ann. N. Y. Acad. Sci.} {\bf 688}, 739
\item  Smith, M. S., Kawano, L. H. \& Malaney, R. A. 1993, {\it ApJS\/}
{\bf 85}, 219
\item Songaila, A., Cowie, L. L., Hogan, C. \& Rugers, M. 1994, {\it
Nature} {\bf 368},
599
\item Spite, M. \& Spite, F. 1982a,  {\it Nature} {\bf 297}, 483
\item Spite, F. \& Spite, M. 1982b, {\it A \& A}, {\bf 115}, 357

\item Steigman, G. 1985, ``Theory and Observational Limits in Cosmology"
(ed. W. R. Stoeger;
Specola Vaticana), p. 149
\item Steigman, G. 1994a, ``Cosmic Lithium: Going Up Or Coming Down?"
(Preprint) OSU-TA-18/94
\item Steigman, G. 1994b, {\it MNRAS\/}, {\bf 269}, L53
\item Steigman, G. \& Tosi, M. 1992, {\it ApJ\/} {\bf 401}, 150
\item Steigman, G. \& Tosi, M. 1994, ``Generic Evolution of Deuterium and
Helium-3"
(Preprint) OSU-TA-12/94
\item Steigman, G. \& Walker, T. P. 1992, {\it ApJ\/} {\bf 385}, L13
\item Thomas, D., Hata, N., Scherrer, R., Steigman, G. \& Walker, T. 1994,
In Preparation
\item Thorburn, J. A. 1994, {\it ApJ\/} {\bf 421}, 318
\item Walker, T. P., Steigman, G., Schramm, D. N., Olive, K. A. \& Kang,
H.-S, 1991, {\it
ApJ\/} {\bf 376}, 51 (WSSOK)
\item White, R. E. 1986, {\it ApJ\/} {\bf 307}, 777
\item White, D. A., Fabian, A. C., Allen, S. W., Edge, A. C., Crawford, C.
S., Johnstone, R.
M., Stewart, G. C., \& Voges, W. 1994, {\it MNRAS\/} {\bf 269}, 598
\item White, D. A. \& Fabian, A. 1994 Preprint (Submitted to MNRAS)
\item White, S. D. M., Navarro, J.
F., Evrard, A. E.
\& Frenk, C. S. 1993, {\it Nature\/} {\bf 366}, 429
\item Yang, J., Turner, M. S., Steigman, G., Schramm, D. N. \& Olive, K. A.
1984, {\it
ApJ\/} {\bf 281}, 493 (YTSSO)\\
\end{itemize}

\end{document}